\documentclass{acm_proc_article-sp}
\usepackage{graphicx} 
\usepackage{caption}
\usepackage{subcaption}
\usepackage{multicol}
\usepackage[procnames]{listings}
\usepackage{tabularx}
\usepackage{color}
\usepackage{listings}  
\usepackage{tabularx}
\definecolor{mygreen}{rgb}{0,0.6,0}
\definecolor{gray}{rgb}{0.4,0.4,0.4}
\definecolor{darkblue}{rgb}{0.0,0.0,0.6}
\definecolor{purple}{rgb}{0.2,0.1,0.3}
\definecolor{background}{rgb}{0.95,0.95,0.95}
\definecolor{myblue}{RGB}{20,105,176}
\lstdefinestyle{styOMP2HMPP}{
  language=C++,
  basicstyle=\tiny,
  backgroundcolor=\color{background},
  numbers=left,
  directivestyle=\textbf,
  keywordstyle=[1]\color{mygreen},
  keywordstyle=[2]\color{purple}\textbf,
  keywordstyle=[3]\color{darkblue}\textbf,
  keywordstyle=[4]\color{mygreen}\textbf,
  identifierstyle=\color{black},
  commentstyle=\color{gray},
  stringstyle=\color{myblue},
  moredirectives={omp, hmpp,  target, CUDA, synchronize, delegatedstore, advancedload, parallel,for, codelet, callsite, private, shared, args, addr,  asynchronous, hmppcg, gridify, release, asynchronous, reduction}, 
  keywords=[1]{myTable,myTableOut, diffsum},
  morekeywords=[2]{in, inout,io}, 
  morekeywords=[3]{group0_46, group, mapbyname, noupdate, true}, 
  morekeywords=[4]{diffsum , diffsum_reduced, reduce}
  deletekeywords={for,int,double, return, true},
  emph={check},
  emphstyle=\color{darkblue}\textbf,
  literate=
            *{\{}{{{\color{myblue}{\{}}}}{1}
            {\}}{{{\color{myblue}{\}}}}}{1}
            {[}{{{\color{myblue}{[}}}}{1}
            {]}{{{\color{myblue}{]}}}}{1},
   escapeinside=!!
}

\lstdefinelanguage{omp2hmpp}
{
 style=styOMP2HMPP,
}

\usepackage[ruled]{algorithm2e}
\SetAlFnt{\algofont}
\SetAlCapFnt{\algofont}
\SetAlCapNameFnt{\algofont}
\SetAlCapHSkip{0pt}
\IncMargin{-\parindent}
\renewcommand{\algorithmcfname}{ALGORITHM}
\newcommand{\includecode}[2][c]{\lstinputlisting[escapechar=]{#2}}

\begin{document}

\title{OMP2HMPP: HMPP Source Code Generation from Programs with Pragma Extensions}

%
%
%
%
%

\numberofauthors{3} 
%
\author{
%
%
\alignauthor
Albert Sa\`{a}-Garriga\\
 \affaddr{Universitat Auton\`{o}ma de Barcelona}\\
 \affaddr{Edifici Q,Campus de la UAB}\\
 \affaddr{Bellaterra, Spain}\\
 \email{albert.saa@uab.cat}
\alignauthor David Castells-Rufas\\
 \affaddr{Universitat Auton\`{o}ma de Barcelona}\\
 \affaddr{Edifici Q,Campus de la UAB}\\
 \affaddr{Bellaterra, Spain}\\
 \email{david.castells@uab.cat}
\alignauthor Jordi Carrabina\\
 \affaddr{Universitat Auton\`{o}ma de Barcelona}\\
 \affaddr{Edifici Q,Campus de la UAB}\\
 \affaddr{Bellaterra, Spain}\\
 \email{jordi.carrabina@uab.cat}
\and
}
\date{29 November 2013}

\maketitle
\begin{abstract}

High-performance computing are based more and more in heterogeneous architectures and GPGPUs have become one of the main integrated blocks in these, as the recently emerged Mali GPU in embedded systems or the NVIDIA GPUs in HPC servers. In both GPGPUs, programming could become a hurdle that can limit their adoption, since the programmer has to learn the hardware capabilities and the language to work with these. 

We present OMP2HMPP, a tool that, automatically translates a high-level C source code(OpenMP) code into HMPP. 
The generated version rarely will differs from a hand-coded HMPP version, and will provide an important speedup, near 113$\times$, that could be later improved by hand-coded CUDA. The generated code could be transported either to HPC servers and to embedded GPUs, due to the commonalities between them.

\end{abstract}



\section{Introduction}

GPGPUs are potentially useful for speed up applications and are one of the main integrated blocks on heterogeneous platforms, an example of these could be the recently emerged Mali GPU which is used embedded systems or the more studied NVIDIA GPUs, which are present on all of the top ten server of the November 2013 list~\cite{Top500}. Although Compute Unified Device Architecture (CUDA)~\cite{CUDA} programming model from NVIDIA, RapidMind~\cite{Rapid}, PeakStream~\cite{Peak}and CTM~\cite{ATI}have made the use of GPUs, for general propose programming easier and efficient. Some graphically rich applications are initially developed for HPC servers and later ported to embedded or mobile platforms. The reduction in available resources means that the application is unlikely to work at the same performance level as it does on the desktop platform. However, due to the commonalities between them a code generated to work with HPC servers will be portable to embedded system. Nevertheless, in both cases the implementation of a program that has to work with GPGPUs will be complex and error-prone due to the programming complexity and the language paradigms. The proposed tool(OMP2HMPP), simplify this task freeing the programmer to learn a new language and to implement the program to work with GPUs. 

High-performance computing(HPC) community has been a\-live for a long time, usually working with a couple of standards: MPI and, the chosen to be the source code input for our tool, OpenMP. OMP2HMPP use of OpenMP directive-based code as input since that facilities the understanding of the source code blocks that can be paralyzed. These blocks will be transformed to work in GPUs using HMPP~\cite{OpenHMPP,dolbeau2007hmpp} set of directives, which are meta-information added in the application source code that do not change the semantic of the original code and blinds the complexity of final architecture to the user. They address the remote execution (RPC) of functions or regions of code on GPUs and many-core accelerators as well as the transfer of data to and from the target. OMP2HMPP uses these directives to define and call GPU kernels, and minimize the use of the directives related to data transfers between CPU and GPU. OMP2HMPP is able to analyze the OpenMP blocks before to transform these and to determine the scope and the aliveness of each of variables used in that block. Finally, OMP2HMPP offers to the programmer a translated version of the original input code which could be able to work in GPUs.

OMP2HMPP is Source to Source compiler (S2S) based on  BSC's Mercurium framework\cite{Mercurium}. Mercurium~\cite{balart2004nanos} is a source-to-source compilation infrastructure aimed at fast prototyping and supports C and C++ languages and is mainly used in Nanos environment to implement OpenMP but since it is quite extensible it has been used to implement other programming models or compiler transformations. This framework is used in order to implement our S2S transformation phases, providing us with the Abstract Syntax Tree(AST) as an easy access to the table of symbols. This information is analyzed through OMP2HMPP tool to parse and translate the original problem to an optimum version of HMPP.

\subsection{HMPP Directives}

The proposed tool is able to use the combination of the following HMPP directives:
\begin{itemize}
\item \textbf{Callsite: }Specifies the use of a codelet at a given point in the program. Related data transfers and synchronization points that are inserted elsewhere in the application have to use the same label.
\item \textbf{Codelet: } Specifies that a version of the function following must be optimized for a given hardware.
\item \textbf{Group: } Allows the declaration of a group of codelets.
\item \textbf{Advanced Load: } Upload data before the execution of the codelet.
\item \textbf{Delegate Store: } The opposite of the advancedload directive in the sense that it downloads output data from the HWA to the host.
\item \textbf{Synchronize: } Specifies to wait until the completion of an asynchronous callsite execution.
\item \textbf{Release: }  Specifies when to release the HWA for a group or a stand-alone codelet .
\item \textbf{No Update: } This property specifies that the data is already available on the HWA and so that no transfer is needed. When this property is set, no transfer is done on the considered argument.
\item \textbf{Target: }Specifies one or more targets for which the codelet must be generated. It means that according to the target specified, if the corresponding hardware is available AND the codelet implementation for this hardware is also available, this one will be executed. Otherwise, the next target specified in the list will be tried. OMP2HMPP always use CUDA since we will test it in a server without OpenCL support.
\end{itemize}

With these directives OMP2HMPP is able to create a version that in the most of the cases will be equal to a HMPP hand-coded version of the original problem.

\subsection{Related Work}

In the recent years, many source-to-source compiler alternatives to the dominant GPU programming models(CUDA~\cite{CUDA} and OpenCL~\cite{opencl08}) have been proposed to overcome the GP\-GPU programing complexity. The more similar alternatives to the tool proposed in this paper are presented to do a brief comparative of these. In contrast to OMP2HMPP, the following explained methods, would obtain a direct transformation to CUDA language, not to HMPP, which means that the CUDA programming complexity is directly exposed to the final user. 

In one hand there are proposals that extend in one way or another current programming standards such a C/C++, OpenMP, etc. to trivialize this task. In the other hand, there are proposals that does not need any previous language extensions to transform the code to work and transform the code directly from CPU to GPUs.

One of the examples that includes language extensions is proposed in~\cite{openacc11std}. CAPS, CRAY, NVIDIA and PGI, which are members of the OpenMP Language Committee published OpenACC in November 2011, an standard for this directive-based programming that contribute to the specification of OpenMP for accelerators. In the same way, but not with the same consensus that OpenACC, in~\cite{lee2009openmp}, a programming interface called OpenMPC is presented. This paper shows and extensive analysis of the actual state of the art in OpenMP to CUDA source-to-source compilers and CUDA optimizers. OpenMPC provides an abstraction of the complex of CUDA programming model and develops an automatic with a user-assisted tuning system. However, OpenMPC and OpenACC require time to understand the new proposed directives, and to manually optimize the data transfer between GPU and GPU. In contradistinction of both cases, OMP2HMPP just add one new OpenMP directive and the programmer forgets to deal with new languages and the optimization of these. 
Another option is hiCUDA directive-based language~\cite{han2009hi}, which is a set of directives for CUDA computation and data attributes in a sequential program. Nevertheless, hiCUDA has the same programming paradigm than CUDA; even though it hides the CUDA language syntax, the complexity of the CUDA programming and memory model is directly exposed to programmers. Moreover, in contrast to OMP2HMPP, hiCUDA does not provide any transfer optimization. 
Finally~\cite{li2011openmp} and~\cite{ayguade2010extending}, propose an OpenMP compiler for hybrid CPU/GPU computing architecture. In this papers they propose to add a directive to OpenMP in order to choose where the OpenMP block must be executed(CPU/GPU). The process is full blinded to the programmer and is a direct translation to CUDA. The main differences with OMP2\-HMPP does not provide any transfer optimization.

There are less proposals that tries to do direct transformation from C/C++ to CUDA without need any new language extension to transform the code. One of this cases is Par4All~\cite{amini2012par4all}. This tool is able to transform codes originally wrote in C or Fortran to OpenMP, CUDA or OpenCL. Par4All transform C/C++ source code and add OpenMP directives where the program thinks that can be useful. This transformation allows the transformation of the created Open\-MP blocks to GPGPUs kernels by transforming the OpenMP directives to CUDA language. However, the transformation that this program done for CUDA have not take in account the kernel data-flow context and this is not optimum in data-transferences.

For source-to-source compiler infrastructure, there are many possible solutions as LLVM~\cite{lattner2004llvm}, PIPS~\cite{ancourt1996pips}, Cetus~\cite{dave2009cetus}, ROSE~\cite{Quinlan00} and the used Mercurium~\cite{balart2004nanos}.

\section{OMP2HMPP Compiler}
\label{sec:compiler}

\begin{figure}[!ht]
\includegraphics[width=\columnwidth]{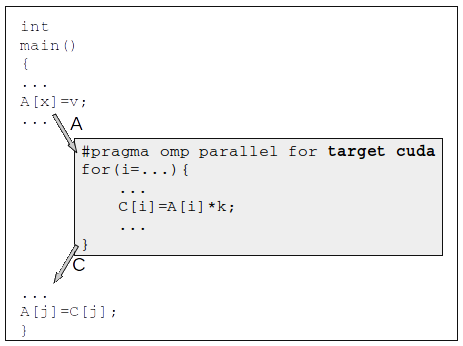}
\caption{Context Example}
\label{fig:toy}
\end{figure}

The main objective of OMP2HMPP is to transform OpenMP blocks into HMPP kernels and their calls. In order to determine the OpenMP blocks that have to be transformed OMP2HMPP use the directives proposed in~\cite{ayguade2009proposal}, as is illustrated in Figure~\ref{fig:toy}. For each block, OMP2HMPP made an outline to create HMPP codelets (GPU kernels) and, at the same time, extract information from the transformed code to determine if the parameters of the created function are used just as input or are updated inside this and therefore, are output. We shown an example of that kernel creation in Table~\ref{src:r} (line 19) for the OpenMP block shown in Table~\ref{src:l} (lines 24-30). 

After that transformation, OMP2HMPP analyze the context where the codelet is invoked, and do an accurate contextual analysis of the AST for each of the variables needed in the created codelet, to reduce the data transfers between CPU and GPU. OMP2HMPP is able to determine where is computed(CPU/GPU) the next or last access(read/write). OMP2\-HMPP search all the assignment (=) or assignment operation expressions (+=,-=,/=,*=) and divide them in two expressions that contain the variables as operands.

\begin{figure}[!ht]
\centering
        \begin{subfigure}[b]{\columnwidth}
                \includegraphics[width=\columnwidth]{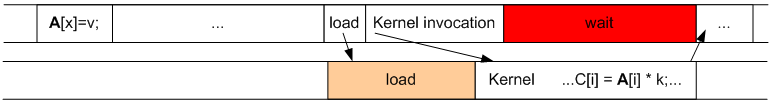}
                \caption{ }
                \label{fig:advloadNO}
        \end{subfigure}
        \begin{subfigure}[b]{\columnwidth}
               \includegraphics[width=\columnwidth]{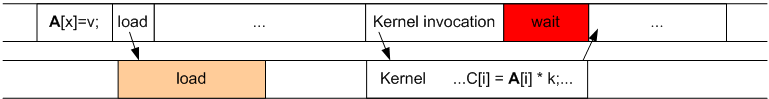}
               \caption{ }
                \label{fig:advloadYES}
        \end{subfigure}
        \caption{Advanced Load Directive Optimization. a) Variables are loaded when kernel is invoked. b) Variables are loaded as near as possible of the last CPU write.}\label{fig:advload}
\end{figure}

\begin{figure}[!ht]
\centering
        \begin{subfigure}[b]{\columnwidth}
                \includegraphics[width=\columnwidth]{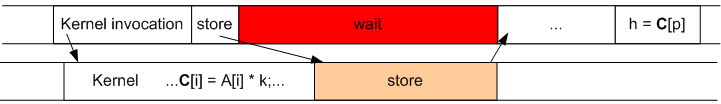}
                \caption{ }
                \label{fig:delstoreNO}
        \end{subfigure}
        \begin{subfigure}[b]{\columnwidth}
                \includegraphics[width=\columnwidth]{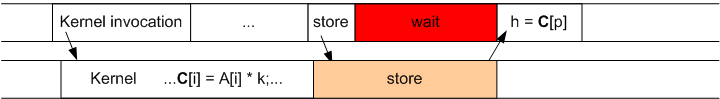}
                \caption{ }
                \label{fig:delstoreYES}
        \end{subfigure}
        \caption{Delegate Store Directive Optimization. a) Variables are downloaded when kernel finish b) Variables are download as far as possible of the kernel finish, next to the first CPU read.}\label{fig:delstore}
\end{figure}

OMP2HMPP use the aforementioned information to choose the best use of HMPP directives that minimize the number of data transfers. For a simple problem, as the proposed in Figure~\ref{fig:toy}, where there is a OpenMP block to transform, OMP2HMPP will analyze the dependences of all the variables inside the kernel (\textit{A} and \textit{C}) and process the extracted information. 
In Figure~\ref{fig:toy}, OMP2HMPP has two variables to analyze \textit{A} and \textit{C}. In the case of \textit{A}, \textit{A} has to be uploaded to GPU, but is not necessary to download after the kernel call because there is no read of that variable before the code finish. In contradistinction, in the case of variable \textit{C}, \textit{C} has to be downloaded from GPU to CPU, but there is no need to upload that to GPU since the kernel do not do a read of \textit{C} inside. With that information, OMP2HMPP will use an \textit{advancedload} in the case of \textit{A} and will put that directive as close as possible to the last write expression, to optimize the data transfer and improve the performance of the generated code as is shown in Figure~\ref{fig:advloadYES}. In the case of \textit{C}, OMP2HMPP will put a \textit{delegatestore} directive, as far as possible of the kernel call, and that will increase the performance of the generated code, as is shown in Figure~\ref{fig:delstoreYES}. Figures ~\ref{fig:advloadNO} and ~\ref{fig:delstoreNO} illustrate the use of a bad transfer policy in the same problems.

\begin{figure}[htb]
\includegraphics[width=\columnwidth]{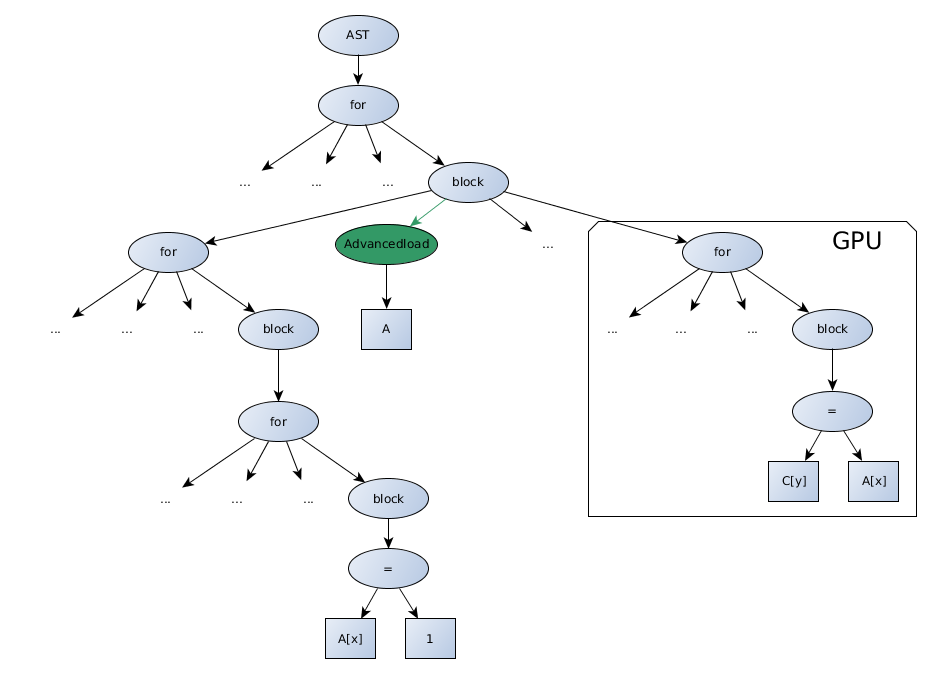}
\caption{Data transfer in Loops Example}
\label{fig:loop}
\end{figure}

\begin{figure}[htb]
\includegraphics[width=\columnwidth]{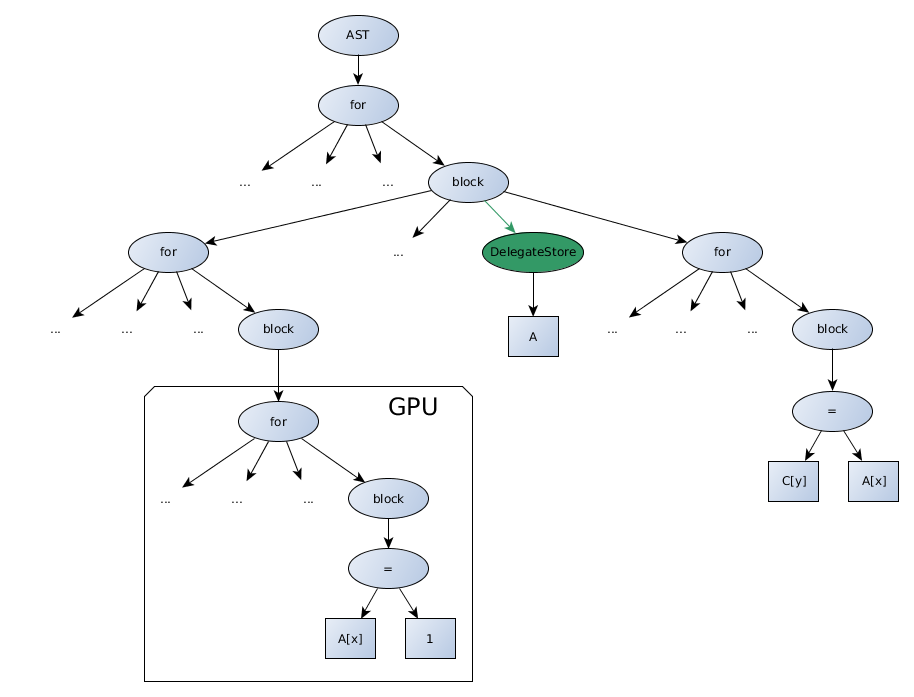}
\caption{Data transfer in Loops Example}
\label{fig:loop2}
\end{figure}

Moreover, OMP2HMPP can deal in context situations in which the source code contains nested loops. OMP2\-HMPP determine if an operation over a variable is made inside a loop and adapt the data transfer to the proper context situation. We illustrate an example of a possible context situations in Figures~\ref{fig:loop} and ~\ref{fig:loop2}. In the first figure, when OMP2\-HMPP wants to compute the loop in GPU, has to load before the values of variable \textit{A}, which is needed to calculate the value of \textit{C}. Since the last write in CPU of \textit{A} is inside a loop with a different nested level than the GPU block, OMP2HMPP has to backtrack the nesting of loops in which is located the last write of \textit{A}, to find the block shared by both loops. Then, as in the problem shown in Figure~\ref{fig:toy}, OMP2HMPP optimize the load of \textit{A} putting the \textit{advancedload} directive as close as possible after the loop finish. We could change the same problem changing the block that is computed in GPU, as is shown in Figure~\ref{fig:loop2}. In this figure, the result of the GPU kernel is needed in CPU to compute \textit{C}, but is not in the same loop level. In that case, the optimum way to put the \textit{delegatestore} directive, will be just before the start of the nested loops where the computation of \textit{C} is located.

\lstset{language=omp2hmpp} 
\begin{table}[ht!]
\begin{tabularx}{\columnwidth}{|X|}
  \hline
  \includecode{src/tab7left.c} \\
  \hline
\end{tabularx}
\caption{OMP2HMPP Original Code Example}
\label{src:l}
\end{table}

\begin{table*}[!ht]
\begin{tabularx}{\textwidth}{|X|}
  \hline
  \includecode{src/tab7right.c} \\
  \hline
\end{tabularx}
\caption{OMP2HMPP Generated Code Example}
\label{src:r}
\end{table*}

\section{Results}
\label{sec:res}

We shown in Table~\ref{src:r} an example of the resulting code for the problem illustrated in Table~\ref{src:l}(3MM). In these figures there is exemplified a real case of use of the OMP2HMPP(explained in Section~\ref{sec:compiler}), we show that OMP2HMPP create a callsite and codelet for each OpenMP block, outlining the OpenMP block into a GPU kernel function, with the syntax requirements of HMPP language. At the same time, OMP2HMPP create a group to share variables between the created kernels using the combination of \textit{group}, \textit{mapbyname} and \textit{noupdate} directives to avoid the unnecessary upload/download(marked in blue in Table~\ref{src:r}). Finally, OMP2HMPP uses asynchronous GPU process of the first kernels since the calculated variables are not needed until the last kernel is invoked.

Furthermore, Table~\ref{src:r} also show the use of the context information and how OMP2HMPP deals with it, as was explained in Section~\ref{sec:compiler}. OMP2HMPP determine the kind of access of each variable needed in the created kernel, as is illustrated in the transformation of the first OpenMP block in Table~\ref{src:l}, where the variable \textit{E} has not been wrote before the block and therefore a load is not necessary. OMP2HMPP also analyze the host where a variable is used (CPU/GPU) as the shown in Table~\ref{src:l} with the variables needed for kernel \textit{\_instr\_for0\_ol\_32\_main}. These variables are all contained and updated in GPU and OMP2HMPP avoids to reload them. Finally, we exemplify the understand of loop context situation and that OMP2HMPP can identify in which loop is the use of an analyzed variable in the case of variable \textit{A} (line 30 in Table~\ref{src:r}), which is needed in HMPP kernel \textit{\_instr\_for0\_ol\_28\_main} (line 53 in Table~\ref{src:r}) and there is no CPU write instructions for this variable between these lines. For that reason, OMP2\-HMPP put \textit{advancedload} instruction after the last variable \textit{A} assignment, but since this assignment is made inside a loop, OMP2HMPP postpone the load after the loop finish and add \textit{advancedload} instruction in line 39.
\begin{figure}[!ht]
\centering
        \begin{subfigure}[b]{\columnwidth}
                \includegraphics[width=\textwidth]{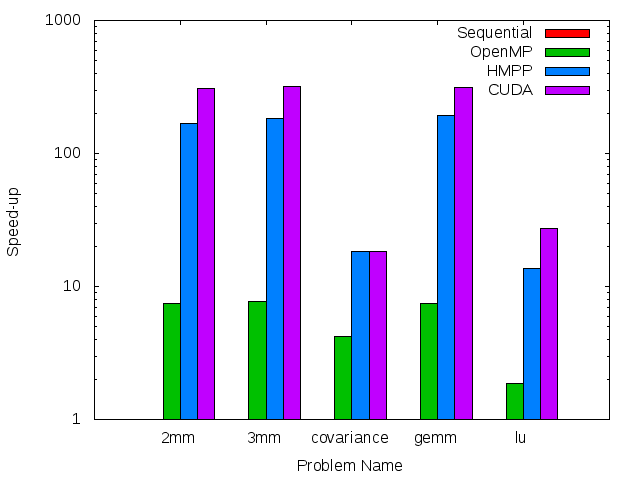}
                \caption{B505 blade(1)}
                \label{fig:a}
        \end{subfigure}
        \begin{subfigure}[b]{\columnwidth}
                \includegraphics[width=\textwidth]{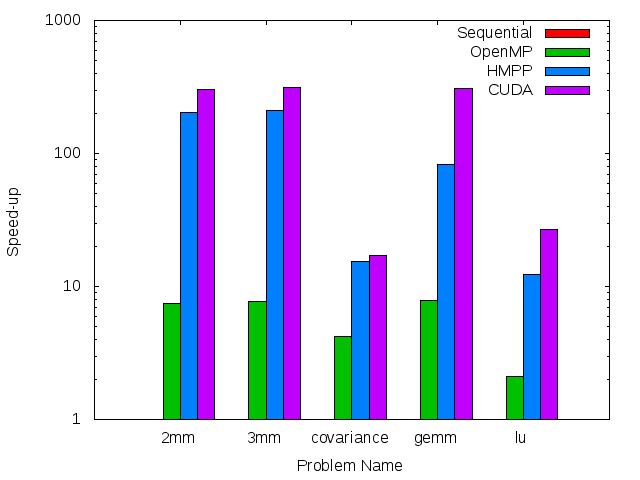}
                \caption{B505 blade(2)}
                \label{fig:b}
        \end{subfigure}
        \begin{subfigure}[b]{\columnwidth}
                \includegraphics[width=\textwidth]{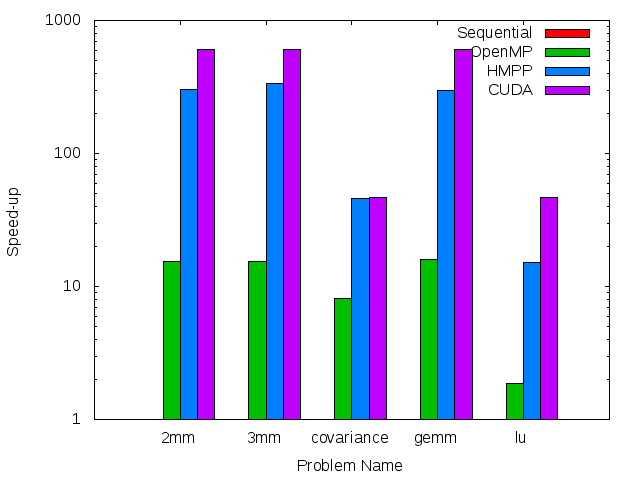}
                \caption{B515 Blade}
                \label{fig:c}
        \end{subfigure}
        \caption{Speedup compared with the sequential version}\label{fig:results}
\end{figure}

\begin{center}
  \begin{table}[!ht]
  \begin{tabularx}{\columnwidth}{| X | X | X | X | }
    \hline
      & B505 blade(1) & B505 blade(2) & B515 blade\\ \hline  \hline
     Num. Processors & 2 & 2 & 2 \\ \hline
     Processor & E5640 & E5640 & E5-2400 \\ \hline
     Memory & 24 Gb & 24 Gb & 192Gb \\ \hline
     GPU & Nvidia Tesla M2050 & Nvidia Tesla C2075 & Nvidia Tesla K20 \\ 
    \hline
  \end{tabularx}
  \end{table}
  \label{tab:arch}
\end{center}

To have an performance of the codes generated by OMP2HMPP, we created a set of codes extracted from the Polybench~\cite{pouchet2012polybench} benchmark and then, we compare the execution of these with the original OpenMP version, with a hand-coded CUDA version and with a sequential version of the same problem. In Figures~\ref{fig:a}~\ref{fig:b}, and ~\ref{fig:c} we show for each problem the speed-up comparison for the architectures showed in Table~\ref{tab:arch}. In these figures, we show that OMP2HMPP made a good transformation for the originally OpenMP code, and obtain an average speedup of 113$\times$. This speedup is less that the obtained in the execution of CUDA hand-coded code of the same problem in the most of the problems with an average speedup of 1.7$\times$, but is quite similar in the covariance problem. Moreover, the average speedup obtained when we compare the generated code to the original OpenMP version is over 31$\times$, which is a great gain in performance for an programmer that do not need any knowledge in GPGPU programming.

\section{Conclusions}

The programmer can use OMP2HMPP source to source compiler and avoid to expend time learning the meaning of HMPP directives or another GPUs language. The tested problems from Polybench benchmark obtains an average speedup of 113$\times$ compared to the sequential version and an average speedup over 31$\times$ compared to the OpenMP version, since OMP2HMPP gives a solution that rarely differ from the best HMPP hand-coded version. The generated version could be useful for more experimented users that want to have, without effort, a GPGPU code that will define an optimization starting point for their problems and then, implement a code that could get, for the tested problems, an speedup near 1.7$\times$ compared with the version generated by OMP2HMPP, as is showed in Section~\ref{sec:res}. 

In future versions of the OMP2HMPP tool, it could be interesting to study the use of Par4All tool to have a complete automatic transformation from a sequential C/C++ source code to an HMPP parallelized version, taking the output OpenMP transformed version of this program as input of our proposal. Moreover, since OMP2HMPP is thought to work in heterogeneous architectures will be interesting to do an exploration of the possibilities to combine CPU and GPGPUs in the same problem. Nevertheless, if we want to use the generated code in in mobile or embedded systems it is important to be aware of power consumption in future versions, and to study the energy/time trade-off of the generated version.

\section{Acknowledgments}

This work was partly supported by the European cooperative ITEA2 projects 09011 H4H and 10021 MANY, the CATRENE project CA112 HARP, the Spanish Ministerio de Econom\'{i}a y Competitividad project IPT-2012-0847-430000, the Spanish Ministerio de Industria, Turismo y Comercio projects and TSI-020100-2010-1036, TSI-020400-2010-120. The authors thank BULL SAS and CAPS Entreprise for their support. 
%
\bibliographystyle{abbrv}
\bibliography{sigproc}

\end{document}